\documentclass[twocolumn, 10pt, journal]{IEEEtran}

\IEEEoverridecommandlockouts
\usepackage{lineno}
\usepackage[font=footnotesize]{caption}  
\usepackage{subcaption}

\usepackage{amsmath}
\usepackage{cite}
\usepackage{float}
\usepackage{lettrine}
\usepackage{soul}
\usepackage{color}
\usepackage[table]{xcolor}
\usepackage{url}
\usepackage{diagbox}
\usepackage{multirow}
\usepackage{array}
\usepackage{bm}
\usepackage{physics}
\usepackage[overload]{empheq}

\usepackage{amsmath}
\usepackage{graphicx}
\usepackage{booktabs} 
\usepackage{times}
\usepackage{fancyhdr,amsmath,amssymb}
\usepackage[ruled,vlined]{algorithm2e}
\include{pythonlisting}
\usepackage{amsmath, amssymb}
\usepackage{makecell}
\usepackage{multirow}
\usepackage{enumitem}
\usepackage{xcolor}
\usepackage{soul}
\usepackage[ruled]{algorithm2e}
\sethlcolor{yellow}

\newcolumntype{M}[1]{>{\centering\arraybackslash}m{#1}}
\newcolumntype{N}[1]{>{\raggedright\arraybackslash}m{#1}}
\newcommand{\eat}[1]

\usepackage{tikz}

\usepackage[usestackEOL]{stackengine}

\stackMath

\usepackage{booktabs}

\usepackage{enumitem}

\usepackage{amsmath}
\usepackage{graphicx}
\usepackage{booktabs} 
\usepackage{times}
\usepackage{fancyhdr,amsmath,amssymb}
\usepackage[ruled,vlined]{algorithm2e}
\include{pythonlisting}
\usepackage{amsmath, amssymb}
\usepackage{makecell}
\usepackage{multirow}
\usepackage{enumitem}
\usepackage{xcolor}
\usepackage{soul}
\usepackage[ruled]{algorithm2e}
\sethlcolor{yellow}
\usepackage{tikz}

\usepackage[usestackEOL]{stackengine}
\stackMath
\usepackage{overpic}

\usepackage{booktabs}

\usepackage{enumitem}
\usepackage{threeparttable}
\usepackage{tabularx}
\newcommand{\PreserveBackslash}[1]{\let\temp=\\#1\let\\=\temp}
\newcolumntype{C}[1]{>{\PreserveBackslash\centering}p{#1}}
\newcolumntype{R}[1]{>{\PreserveBackslash\raggedleft}p{#1}}
\newcolumntype{L}[1]{>{\PreserveBackslash\raggedright}p{#1}}

\newcommand{\peng}[1]{\ifthenelse{\boolean{showcomments}}
{ \textcolor{magenta}{(Peng says:  #1)}}{}}
\newcommand{\yue}[1]{\ifthenelse{\boolean{showcomments}}
{ \textcolor{red}{(Yue says:  #1)}}{}}
\newcommand{\fei}[1]{\ifthenelse{\boolean{showcomments}}
{ \textcolor{blue}{(Fei says:  #1)}}{}}
 
\graphicspath{ {fig/} }

\begin{document}

\title{Noise-Resilient Quantum Power Flow}

\author{Fei~Feng,~\IEEEmembership{Student Member,~IEEE}, Yifan~Zhou,~\IEEEmembership{Member,~IEEE} and Peng~Zhang,~\IEEEmembership{Senior Member,~IEEE}

\thanks{This material is based upon work supported  in part  by the U.S. Department of Energy's Office of Energy Efficiency and Renewable Energy (EERE) under the Solar Energy Technologies Office Award Number 38456, and in part by the National Science Foundation under Grant  No. OIA-2134840. Any opinions, findings, and conclusions, or recommendations expressed herein do not necessarily represent the views of the Department of Energy, the National Science Foundation, or the United States Government. }
\thanks{The authors are with the Department of Electrical and Computer Engineering, Stony Brook University, Stony Brook, NY 11794-2350, USA (e-mail: yifan.zhou.1@stonybrook.edu).}
}

\markboth{}
{Shell \MakeLowercase{\textit{ et al.}}:  Bare Demo of IEEEtran.cls for IEEE Journals}

\maketitle

\begin{abstract}

Quantum power flow (QPF) provides inspiring directions for tackling power flow's computational burdens leveraging quantum computing. However, existing QPF methods are mainly based on noise-sensitive quantum algorithms, whose practical utilization is significantly hindered by the limited capability of today's noisy-intermediate-scale quantum (NISQ) devices. 
This paper devises a NISQ-QPF algorithm, which enables power flow calculation on noisy quantum computers. 
The main contributions include:
(1) a variational quantum circuit (VQC)-based AC power flow formulation, which enables QPF using short-depth quantum circuits;
(2) noise-resilient QPF solvers based on the variational quantum linear solver (VQLS) and modified fast decoupled power flow;
(3) a practical NISQ-QPF framework for implementable and reliable power flow analysis on noisy quantum machines.
Promising case studies validate the effectiveness and accuracy of NISQ-QPF on IBM's real, noisy quantum devices.

\end{abstract}

\IEEEpeerreviewmaketitle

\section{Introduction}
\IEEEPARstart{P}{ower} flow is the keystone for various modern power system analytics (e.g., stochastic power flow, security screening, and reliability assessment)~\cite{feng2020enhanced, 7177142,feng2022distributed}. Under the deep penetration of renewables, an enormous amount of power flow analyses are needed to quantify the impact of uncertainties~\cite{7185462}. However,  power flow calculation remains to be intractable because the complexities of almost all the classical iterative algorithms scale polynomially with the system size.  

The fast evolution in quantum computing provides a promising direction for developing scalable power flow analytics~\cite{zhou2022quantum,Chen_2019,zhou2021quantum}. 
Unlike classical methods, quantum computing enables using logarithmically-scaled number of qubits to solve linear equations in power flow analysis.
A Harrow-Hassidim-Lloyd (HHL)-based quantum power flow (QPF) is devised to underpin the AC power flow issue through quantum computing~\cite{feng2021quantum}. Although the proof-of-concept is successful, the scalability of the method remains limited. The main obstacle is that HHL generates high-depth quantum circuits even for small-scale power flow problems, which can be significantly crippled by noise. 
Today, noisy-intermediate-scale quantum (NISQ) devices remain to be mainstream, whose capability is restricted by the limited number of qubits and considerable noises. Noise-tolerant quantum devices may not be available in the near future due to the significant error correction overhead and the short coherence time~\cite{nielsen2000quantum,feng2022quantum}. Therefore, QPF methods that are applicable to NISQ devices are in high need.

To bridge the gap, this paper devises a  NISQ-QPF algorithm that allows for practical and noise-resilient power flow analysis on NISQ devices. Our contributions are as follows:
\begin{itemize}[leftmargin=*]
  \item  A variational quantum circuit (VQC)-based QPF formulation is established to enable QPF analysis using shallow-depth quantum circuits.
  \item  A variational quantum linear solver (VQLS)-QPF solver is devised incorporating power flow information embedding and quantum circuit optimization, which enables 
  {{resilient quantum}} power flow iterations under noisy 
  environments. 
  \item Practical VQLS-QPF solvers are devised by modifying the fast decoupled power flow, which overcomes the quantum measurement issues of the general VQLS.
  \item  A NISQ-compatible QPF framework is constructed for reliable QPF implementation on real quantum computers.  
  
\end{itemize}

The remainder of this paper is organized as follows: Section II establishes the variational QPF formulation. Section III develops the NISQ-compatible QPF algorithm. Section IV presents extensive case studies on real IBM quantum devices, followed by the Conclusion in Section V.

\section{ Variational Quantum Power Flow Formulation} \label{sec:model}
An indispensable computation burden of nonlinear power flow algorithms is to solve a set of linear equations iteratively, where the required computational resource scales polynomially with the power system scale. This section establishes a variational QPF formulation, which inherits the exponential scalability of quantum computing and enables the utilization of shallow-depth quantum circuits in QPF calculation.

\subsection{Classical Fast Decoupled Load Flow (FDLF) Formulation}
Given an $N$-bus power system with one slack bus, $N_{pv}$ PV buses, and $N_{pq}$ PQ buses, the fast decoupled load flow (FDLF) models are formulated as~\cite{stott1974fast}:
\begin{equation}\label{eq:FDPFp}
 {\pmb{V}^{-1}\Delta\pmb P}=   {\pmb B}^{'}{{ \pmb{V}}\Delta{\pmb{\theta}}}
\end{equation}
\begin{equation}\label{eq:FDPFq}
  {{\pmb{V}}^{-1}\Delta{\pmb Q}}= {\pmb B}^{''}{ \Delta{\pmb{V}}}
\end{equation}

\noindent Here, ${\pmb B}^{'}\in \mathbb{R}^{(N-1) \times (N-1)}$ and ${\pmb B}^{''}\in \mathbb{R}^{N_{pq} \times N_{pq}} $ are coefficient matrices derived from the admittance matrix; $\Delta{ \pmb V}\in \mathbb{R}^{N_{pq} \times 1}$ and $\Delta{\pmb{\theta}} \in \mathbb{R}^{(N-1) \times 1}$ are the differences of voltage magnitudes and angles, respectively; 
$\Delta{ \pmb P}\in \mathbb{R}^{(N-1) \times 1}$ and $\Delta{\pmb{Q}} \in \mathbb{R}^{N_{pq} \times 1}$ 
denote the active/reactive power mismatches, which is updated as: 
\begin{equation}\label{eq:FDPF}
  \Delta{\pmb {S}}=[\Delta{\pmb{P}} ,\Delta{\pmb Q}]^T=
{\pmb{S}}-{\Bar{\pmb Y}({\pmb{\theta}})} \cdot {\pmb{V}} \circ{\pmb{V}} 
  \end{equation}

\vspace{-5pt}
\subsection{VQC-Based QPF Formulation}


The main idea of NISQ-QPF is to establish two separate VQCs to respectively prepare $\ket{\Delta {\pmb V}}$ and $\ket{{ \pmb{V}}\Delta{\pmb{\theta}}}$ according to \eqref{eq:FDPFp} and \eqref{eq:FDPFq} (here $\lvert \cdot \rangle$ denotes the corresponding quantum state of the original vector\footnote{Taking 
$\Delta {\pmb V}$ as an example, 
$\lvert \Delta {\pmb V} \rangle = \sum\nolimits_{j}\nu_j\lvert j\rangle$, where $\nu_j={ \Delta{V}_j} / {\sqrt{\sum_j ( \Delta{V}_j)^2}}$; $\Delta{V}_j$ denote the $j^{th}$ element of $ \Delta{\pmb{V}}$; $\lvert j\rangle$ is the $j^{th}$ quantum basis.}).

Taking \eqref{eq:FDPFq} as an example, we explain the formulation of the fast decoupled, VQC-based QPF.
A VQC $ {U}_q(\bm w_q) $ specified by a set of classical parameters $\bm{w}_q$ generates $\ket{\Delta \pmb{V}}$ as
$\ket{\Delta \pmb{V}} = {U}_q (\bm w_q)  \ket{\bm{0}}$. In order to optimize a qualified ${U}_q (\bm w_q)$ so that $\ket{\Delta \pmb{V}}$ provides the solution of \eqref{eq:FDPFq}, it is required that ${\pmb B}^{''}\ket{\Delta \pmb{V}}$ (i.e., the left-hand side of \eqref{eq:FDPFq}) is proportional to the normalized quantum state of $\pmb{V}^{-1}\Delta{\pmb Q}$  (i.e., the right-hand side of \eqref{eq:FDPFq}). Mathematically, the following formulation is established:
\begin{equation}\label{eq:QPF}
     \ket{\bm{\Psi}_q} = \frac{{\pmb B}^{''} \lvert \Delta \pmb{V} \rangle}{ \sqrt{  \expval{ {{\pmb B}^{''}}^T  {\pmb B}^{''}}{\Delta \pmb{V} } }} = \lvert {\pmb{V}}^{-1}\Delta{\pmb Q} \rangle
\end{equation}
where $\ket{\bm{\Psi}_q}$ denotes the normalized quantum state of $\pmb{V}^{-1}\Delta{\pmb Q}$.

 For \eqref{eq:FDPFp}, the preparation of $\ket{{ \pmb{V}}\Delta{\pmb{\theta}}}$ can be devised as the same format of \eqref{eq:QPF}:
\begin{equation}\label{eq:QPF normalized P}
     \ket{\bm{\Psi}_p} = \frac{{\pmb B}^{'} \ket{{ \pmb{V}}\Delta{\pmb{\theta}}} }{ \sqrt{  \expval{ {{\pmb B}^{'}}^T  {\pmb B}^{'}}{ { \pmb{V}}\Delta{\pmb{\theta}} } }} = \lvert {\pmb{V}}^{-1}\Delta{\pmb P} \rangle
\end{equation}
where $\ket{{ \pmb{V}}\Delta{\pmb{\theta}}} = {U}_p (\bm w_p)  \ket{\bm{0}}$ is yielded from another VQC $ {U}_p $ specified by parameters $\bm{w}_p$). 

\begin{figure}[t!]
  \centering
  \includegraphics[width=0.5\textwidth]{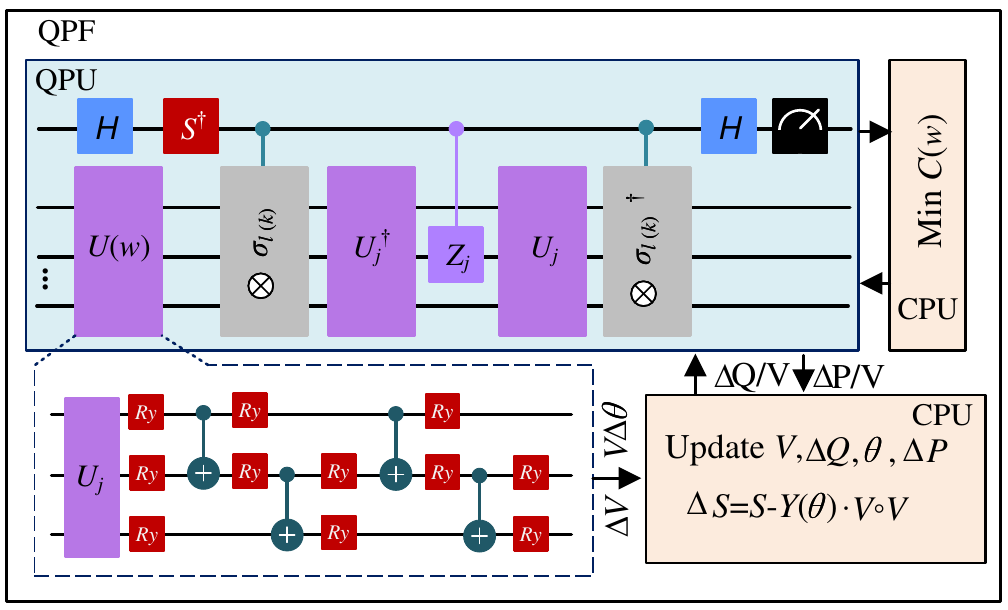}
  \caption{QPF architecture}
  \label{fig:QPFcircuit}
  \vspace{-10pt}
\end{figure}

\section{Noise-Intermediate-Scale Quantum Power Flow Algorithm}
This section develops the NISQ-QPF algorithm. 
A VQLS-based QPF solver is first developed for power flow iterations. Then, practical QPF solvers are devised to address the measurement challenges on real quantum machines.
Finally, practical considerations on circuit design and implementation settings are discussed.

\subsection{VQLS-based QPF Solver}

This subsection establishes a VQLS-based QPF solver to optimize the VQCs for the variational QPF formulated in Section \ref{sec:model}.
Fig.~1 illustrates the architecture of the  VQLS-based QPF algorithm, where a hybrid quantum/classical framework is utilized. Detailed procedures are as follows:

\subsubsection{Power flow information updating}
According to \eqref{eq:FDPFp}-\eqref{eq:FDPF}, the power injections and voltages are updated to obtain $ {\pmb{V}}^{-1}\Delta{\pmb P}$ and $ {\pmb{V}}^{-1}\Delta{\pmb Q}$ in the CPU, which only involves trivial matrix-vector multiplication.

\subsubsection{ VQLS input preparation} 
The quantum processor unit (QPU) reads the power flow vectors ($ {\pmb{V}}^{-1}\Delta{\pmb P}$ and $ {\pmb{V}}^{-1}\Delta{\pmb Q}$) and power flow matrices (${\pmb B}^{'}$ and ${\pmb B}^{''}$) as inputs for the VQLS algorithm to perform linear equation solving in each power flow iteration. 
Correspondingly, the classical matrices/vectors should be translated into the quantum formulation.
To achieve this goal, ${\pmb B}^{'}$ and ${\pmb B}^{''}$ are decomposed into a linear combination of basic Pauli gates~\cite{zhou2022noisy}. $\lvert {\pmb{V}}^{-1}\Delta{\pmb Q} \rangle$ and $\lvert {\pmb{V}}^{-1}\Delta{\pmb P} \rangle$ are decomposed into a basis vector-based formulation so that they can be effectively prepared in the Hilbert space~\cite{zhou2022noisy}. 


\subsubsection{VQC optimization for QPF}
We establish QPF cost functions to quantify the difference between the normalized states in \eqref{eq:QPF} and \eqref{eq:QPF normalized P} (i.e., $\ket{\bm{\Psi}_p}$ and $\ket{\bm{\Psi}_q}$) and power flow vectors (i.e., $\lvert {\pmb{V}}^{-1}\Delta{\pmb P} \rangle$ and $\lvert {\pmb{V}}^{-1}\Delta{\pmb Q} \rangle$):
\begin{equation}\label{eq:qPFcost1}
   \begin{aligned}
 \mathcal{C}_p {=} 1 {-}|\braket{{\pmb{V}}^{-1}\Delta{\pmb P}}{\bm{\Psi}_p}|^2 ~~,~~
 \mathcal{C}_q {=} 1 {-}|\braket{{\pmb{V}}^{-1}\Delta{\pmb Q}}{\bm{\Psi}_q}|^2 
  \end{aligned}
\end{equation}
Values of $\mathcal{C}_p$ and $\mathcal{C}_q$ can be estimated by performing a series of Hadamard tests, which is a standard quantum computation technique \cite{bravo2020variational}.


Then, \eqref{eq:qPFcost1} is minimized to optimize the parameters of the VQCs ${U}_p(\bm w_p)$ and ${U}_q(\bm w_q)$. The subplot in Fig.\ref{fig:QPFcircuit} illustrates the VQC architecture designed for QPF, which is comprised of an initial state encoding layer and several entangling layers~\cite{zhou2022noisy}.

Once the optimization achieves convergence, the maximum overlap can be reached for \eqref{eq:QPF} and \eqref{eq:QPF normalized P}. Then, the possibility of the quantum state $\ket{\Delta \pmb{V}}$ and  $\ket{{ \pmb{V}}\Delta{\pmb{\theta}}}$ can be measured from the optimized VQCs.

The superiority of the fast-decoupled, VQLS-based QPF lies in three points: 
 \begin{itemize}[leftmargin=*]
\item It embeds power flow equations into the Hilbert space, which only uses logarithmically-scaled computational resources (i.e., number of qubits) to address power flow issues. 
\item It employs quantum circuits with shallow-depth structures, which are less crippled by noise and, therefore, more compatible with today's real quantum machines.
 \item It inherits the fixed Jacobian matrix from FDLF so that VQLS only needs to be called at the beginning of the QPF algorithm (rather than at each iteration), which significantly saves the efforts for quantum circuit optimization. 
 \end{itemize}

\vspace{-10pt}
\subsection{Practical NISQ-QPF Algorithm }
The aforementioned VQLS-based QPF solver enables obtaining the quantum states of $\ket{\Delta \pmb{V}}$ and $\ket{{ \pmb{V}}\Delta{\pmb{\theta}}}$ theoretically. However, the measurement on real quantum machines can only provide probabilities on the quantum basis, i.e., ${\Delta {V_j}^2}$ and $({V_j \Delta{{\theta}_j}})^2$  rather than ${\Delta {V_j}}$ and ${V_j \Delta{{\theta}_j}}$  ($\forall j$). 
The main bottleneck lies in the unknown signs of ${\Delta {V_j}}$ and ${V_j \Delta{{\theta}_j}}$.
Here, we devise practical active/reactive-power-related QPF solvers to tackle this challenge.


\subsubsection{Active-power-related QPF solver} 
Rewrite $ \ket{{\pmb{V}}^{-1}\Delta{\pmb P}} $ in its basis states as $\lvert {\pmb{V}}^{-1}\Delta{\pmb P} \rangle = \sum\nolimits_{j}\nu_j\lvert j\rangle$.
Correspondingly, the active power flow solution $\ket{{ \pmb{V}}\Delta{\pmb{\theta}}}$ can also be described as the linear combination of a set of basis solutions  $\ket{{ \pmb{V}}\Delta{\pmb{\theta}}}_j$ ($\forall j$): 
\begin{align}\label{eq:B'basis}
   &  -{\pmb B}^{'}\ket{{ \pmb{V}}\Delta{\pmb{\theta}}}_j=\ket{j} \\
&  \ket{{ \pmb{V}}\Delta{\pmb{\theta}}}=\sum\nolimits_j -\nu_j \ket{{ \pmb{V}}\Delta{\pmb{\theta}}}_j
\end{align}

In the FDLF formulation,  ${\pmb B}^{'} $ is constructed by the reciprocal of branch susceptance~\cite{stott1974fast}:
\begin{align}\label{eq:B'element}
{ B}^{'}_{(g,g)}  = \sum_{s=1, s\neq g}^{N} -\frac{1}{x_{(g,s)}} ~~,~~{ B}^{'}_{(g,m)}  =  \frac{1}{x_{(g,m)}}
\end{align}
where $x_{(g,s)}$ denotes the susceptance of branch $(g,m)$.
Equation \eqref{eq:B'element} shows ${ B}^{'}$ is a diagonally dominant matrix. Therefore, it can be proved that for an arbitrary $j$,  every element of $\ket{{ \pmb{V}}\Delta{\pmb{\theta}}}_j$ is non-negative.

\noindent{\textit{\textbf{Proof:}}} 
Assume bus $N$ is the slack bus. Denote $\ket{{ \pmb{V}}\Delta{\pmb{\theta}}}_j = [{ {V}}_1\Delta{{\theta}}_1, { {V}}_2\Delta{{\theta}}_2,\cdots,{ {V}}_g\Delta{{\theta}}_g, { {V}}_{N-1}\Delta{{\theta}}_{N-1}]^T$.
Without loss of generality, denote ${ {V}}_g\Delta{{\theta}}_g$ as the smallest element in ${{ \pmb{V}}\Delta{\pmb{\theta}}}$, we prove ${ {V}}_g\Delta{{\theta}}_g \geqslant 0$ by contradiction. 

Assume ${ {V}}_g\Delta{{\theta}}_g < 0$.
The $g^{th}$ dimension of \eqref{eq:B'basis} can be calculated as:
\begin{equation}\label{eq:CPFp}
  {j}_g=  \sum_{s=1, s\neq g}^{N-1} \frac{1}{x_{(g,s)}} ({ {V}}_g\Delta{{\theta}}_g-{ {V}}_{s}\Delta{{\theta}}_{s}) + \frac{1}{x_{(g,N)}} { {V}}_g\Delta{{\theta}}_g
\end{equation}
\noindent where  $j_g\in \{0,1\}$ denotes the $g^{th}$ dimension of $\ket{j}$.

Since ${ {V}}_g\Delta{{\theta}}_g$ is the smallest, ${ {V}}_g\Delta{{\theta}}_g \leq { {V}}_{s}\Delta{{\theta}}_{s}$ for arbitrary $s$. Meanwhile, since ${ {V}}_g\Delta{{\theta}}_g < 0$, \eqref{eq:CPFp} yields the following conclusion: 
\begin{equation}\label{ieq:CPFp}
\begin{aligned}
{j}_g & \leq \sum_{s=1, s\neq g}^{N-1} \frac{1}{x_{(g,s)}} ({ {V}}_g\Delta{{\theta}}_g-{ {V}}_g\Delta{{\theta}}_g) + \frac{1}{x_{(g,N)}} { {V}}_g\Delta{{\theta}}_g \\
& =\frac{1}{x_{(g,N)}} { {V}}_g\Delta{{\theta}}_g < 0
\end{aligned}  
\end{equation}

Equation \eqref{ieq:CPFp} is in contradiction to the fact that $j_g\in \{0,1\}$. Therefore, the assumption ${ {V}}_g\Delta{{\theta}}_g < 0$ does not hold and ${ {V}}_g\Delta{{\theta}}_g$ is proved as non-negative.

Consequently, once a VQC is optimized for \eqref{eq:B'basis}, the quantum state $\ket{{ \pmb{V}}\Delta{\pmb{\theta}}}_j$ (rather than merely the possibility) can be estimated as:
\begin{equation}\label{eq:v final}
    \ket{{ \pmb{V}}\Delta{\pmb{\theta}}}_j = {U}_{pj} (\bm w_{pj}^*)  \ket{\bm{0}} = \sum\nolimits_{k=1}^{2^n} \sqrt{p_k}  \ket{k}
\end{equation}
where $p_k$ denotes possibility of the $k^{th}$ basis state, which can be directly measured from real quantum computers.

\subsubsection{Reactive-power-related QPF solver} 
In the FDLF formulation, ${\pmb B}^{''} $ is constructed from the imaginary part of branch and ground admittance. Different from ${\pmb B}^{'} $, ${\pmb B}^{''} $ does not hold the diagonal dominance characteristic,  which therefore hinders the non-negative measurement of the basis solution $\ket{ \Delta\pmb{V}}_j$. 

Inspired by the instant updating process~\cite{zimmerman1995comprehensive,feng2020implicit}, a modified $\pmb{Q}-\pmb{V}$ iteration is devised to tackle the challenge:
\begin{equation}\label{eq:FDPFqmodify}
  {{\pmb{V}}^{-1}\Delta{\pmb Q}}= \Tilde{\pmb B}^{''}{ \Delta{\pmb{V}}}+\Tilde{\pmb B}_0^{''}{ \Delta{\pmb{V}_p}}
\end{equation}
\noindent where $\Delta{\pmb{V}_p}$ denotes the differences of voltage magnitude at the previous iteration. Here, we decompose ${\pmb B}^{''} $ into a branch admittance  related $\Tilde{\pmb B}^{''} $ and a ground admittance related $\Tilde{\pmb B}_{0}^{''} $. The detailed elements  of $\Tilde{\pmb B}^{''} $ and $\Tilde{\pmb B}_{0}^{''} $ include:
\begin{align}\label{eq:B''element}
\Tilde{ B}^{''}_{(g,g)}  = \mathbf{Im}(\sum_{s=1, s\neq g}^{N} y_{(g,s)}) ,\Tilde{ B}^{''}_{(g,m)}  =  \mathbf{Im}({ -{y_{(g,m)}}})
\end{align}
\vspace{-10pt}
\begin{align}\label{eq:B0''element}
\Tilde{ B}^{''}_{0(g,g)}  = \mathbf{Im}( y_{(g,0)}) ,\Tilde{ B}^{''}_{0(g,m)}  =  0
\end{align}
\noindent where $y_{(g,m)}$ and $y_{(g,0)}$ denotes admittance values of branch $(g,m)$ and ground, respectively. 

The modified $\pmb{Q}-\pmb{V}$ iteration retains both the contribution of the ground part and the diagonal dominance characteristic of $ \Tilde{\pmb B}^{''}$, which therefore enables non-negative measurement of $\ket{\Delta{\pmb{V}}}$ from \eqref{eq:FDPFqmodify}. Proof of the modified $\pmb{Q}-\pmb{V}$ iteration shares the same way to \eqref{eq:CPFp} and \eqref{ieq:CPFp}.

\subsubsection{NISQ-QPF Algorithm}
Once the quantum states of $\ket{{ \pmb{V}}\Delta{\pmb{\theta}}}$ and $\ket{\Delta \pmb{V}}$ are measured, 
 the corresponding power flow variables  $\pmb{\theta}$, $\pmb{V}$ and $\Delta \pmb S$ can be updated for the next iteration. The QPF iterations continue until the mismatches $\Delta \pmb S$  reach a convergence tolerance of $\xi$.   \textbf{Algorithm}\ref{VQLS-QPF} details the pseudo-code of QPF.

\begin{algorithm}
\SetAlgoLined
  \textbf{Initialize:} $\pmb{\theta}$, $\pmb{V}$, ${\pmb B}^{'}$, $ \Tilde{\pmb B}^{''}$, $ \Tilde{\pmb B}_0^{''}$, $\pmb S$, $\xi$\; 
 \While{$\Delta\pmb S$\(\geq\)$\xi $ }
 { 
   Update: $\Delta\pmb P$, $\Delta\pmb Q$ \textbf{Eq.}~(\ref{eq:FDPF})\;
  \If{$1^{st}$ iteration}{ Decompose: 
   ${\pmb B}^{'}$, $ \Tilde{\pmb B}^{''}$ $\Rightarrow$ ${\pmb B}^{'}$, $ {\pmb B}^{''}$ \;
  Prepare QPF circuit and $\mathcal{C}$ \textbf{Eq.}~\eqref{eq:qPFcost1} \;
    Optimize QPF circuit \;
   }
  Execute: $\lvert {\pmb{V}}^{-1}\Delta{\pmb P} \rangle$$\xrightarrow{QPF}$$\ket{{ \pmb{V}}\Delta{\pmb{\theta}}}$\;
  Execute: $\lvert {\pmb{V}}^{-1}\Delta{\pmb Q}-\Tilde{\pmb B}_0^{''}{ \Delta{\pmb{V}_p}} \rangle$$\xrightarrow{QPF}$ $\ket{\Delta \pmb{V}}$\;
  
  Update: $\pmb{\theta}$, $\pmb{V}$\; 
 }
 \KwResult{$\pmb{\theta}$, $\pmb{V}$ and the branch power flow. }
 \caption{QPF Algorithm}
 \label{VQLS-QPF}
\end{algorithm}
\vspace{-10pt}

\subsubsection{Remarks}
Today's NISQ computers are considerably disturbed by noise. To enhance the performance of the devised VQLS-QPF solvers, proper circuit design and operation settings on quantum machines are highly required.  Here, the following practical factors are emphasized: 


\begin{itemize}[leftmargin=*]
\item \textbf{Number of shots}. Sufficient quantum shots are required for generating  accurate measurements when solving QPF on real quantum devices. 

\item   \textbf{Quantum circuit structure}.  
The real quantum device usually only contains several basis gates and provides limited connectivity. If the designed VQC requires nonexistent quantum gates or connection between unconnected qubits, it will be compiled to an equivalent quantum circuit compatible with the hardware availability. Such a compiling process unavoidably increases the depth of the quantum circuit, which may deteriorate the performance of QPF.

\end{itemize}

\section{Case Study}
\begin{figure}[ht]
  \centering
  \includegraphics[width=0.45\textwidth, height=0.18\textwidth]{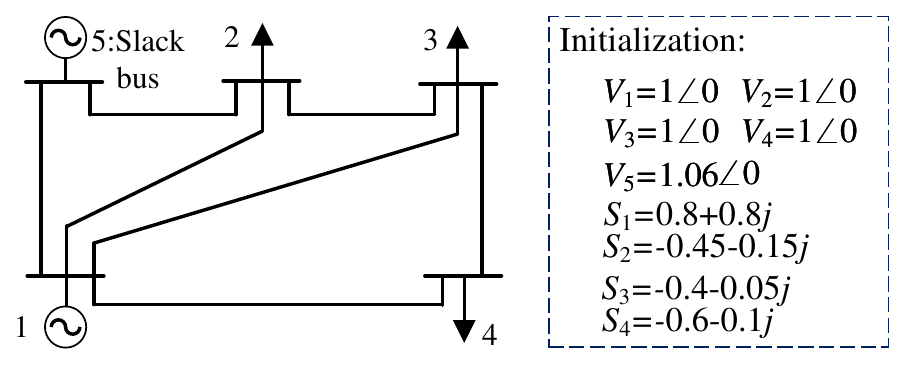}
  \caption{Five-bus system for QPF tests}
  \label{Case5bus}
  \vspace{-10pt}
\end{figure}

This section validates the effectiveness and accuracy of QPF in a typical five-bus test system (see Fig.\ref{Case5bus}) and a nine-bus test system. QPF is implemented on an IBM quantum simulator ({\it{IBMQ\_qasm\_simulator}}) with Qiskit(0.16.0) and a practical quantum machine ({\it{IBMQ\_belem}}). 
\vspace{-10pt}
\subsection{Validity of NISQ-QPF on the Quantum Simulator}
This subsection validates the effectiveness of QPF on an IBM quantum simulator. 

\subsubsection{Validation of the NISQ-QPF process} 
\begin{figure}[!ht]
  \centering
  \includegraphics[width=0.5\textwidth]{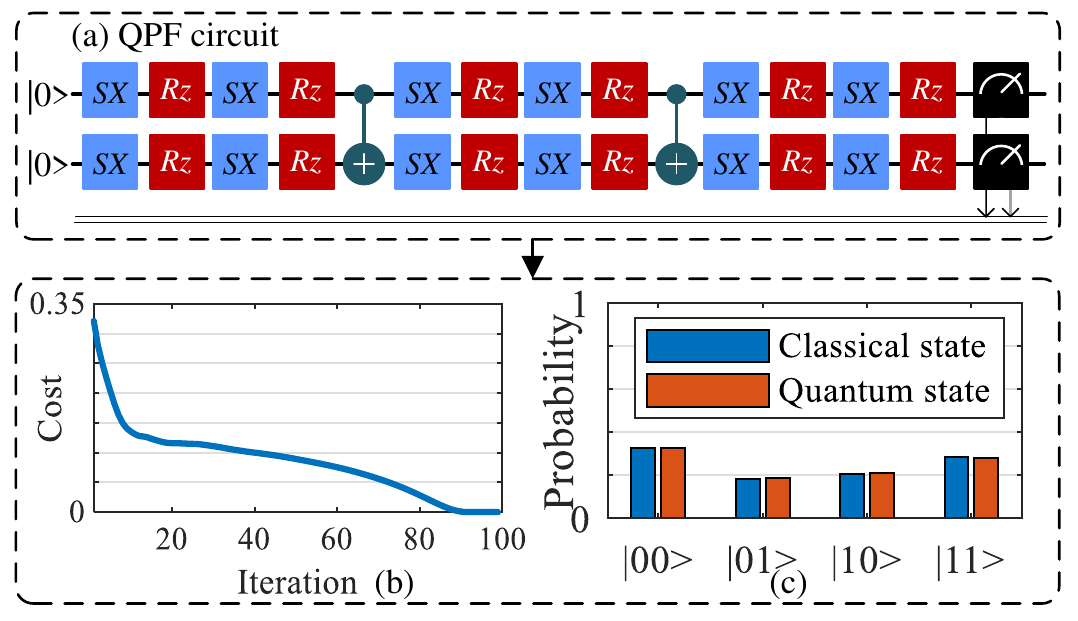}
  \caption{Demonstration of the QPF process in a five-bus test system. (a) Designed QPF circuit; (b) Optimization process of the cost function for a basis state ${\pmb{V}}^{-1}\Delta{\pmb Q} = [1, 0, 0, 0]$; (c) Measurement of quantum state}
  \label{fig:QPFprocess}
  \vspace{-5pt}
\end{figure}

We first exemplify the QPF results on a five-bus test system. 2 qubits are employed to obtain 4 unknown bus voltages. Fig.~\ref{fig:QPFprocess} illustrates the optimized QPF quantum circuit and the measurement results for $\pmb{Q}-\pmb{V}$ iteration under one basis state.
\begin{itemize}[leftmargin=*]
  \item  The compiled, optimized QPF circuit is presented in Fig.~\ref{fig:QPFprocess}(a). It is obvious that the circuit is of shallow depth, which is executable on today's NISQ devices.
  \item Fig.~\ref{fig:QPFprocess}(b) provides the evolution of the cost function . 
    Along the optimization process, the cost function is minimized, which indicates the quantum solution is getting closer to the real solution. 
  \item When the cost function achieves convergence, the quantum states can be measured from the optimized circuit.  Fig.~\ref{fig:QPFprocess}(c) illustrates the perfect match between the measured quantum solutions and the classical power flow solutions, which validates the accuracy of the VQLS-based QPF solver.
  
\end{itemize}

\vspace{-5pt}

\subsubsection{Comparison with classical results} QPF's correctness and convergence are validated by comparing QPF results against those from classical power flow methods. 

\begin{figure}[!ht]
  \centering
  \includegraphics[width=0.45\textwidth]{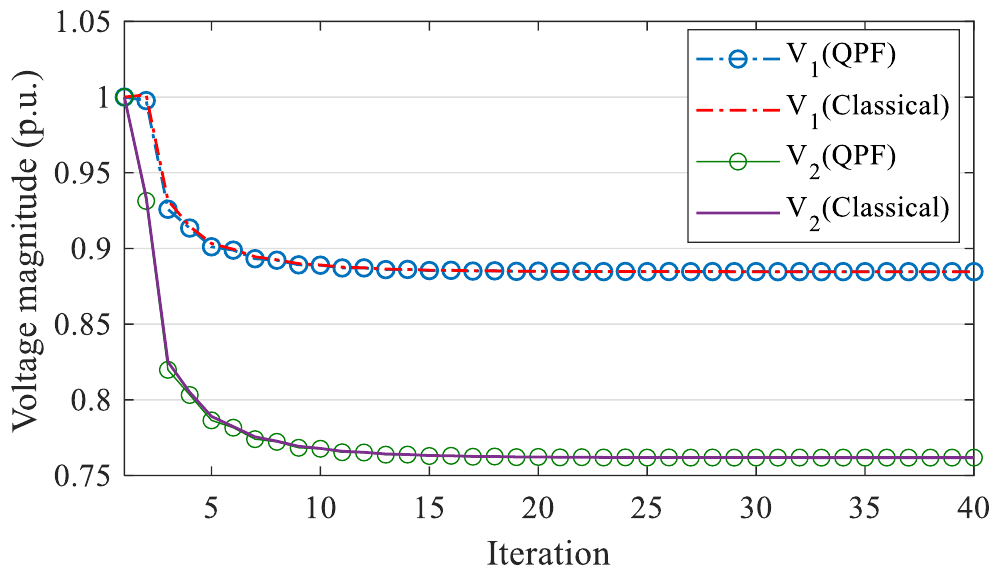}
  \caption{Voltage profiles under the stressed condition }
  \label{fig:V_heavy}
  \vspace{-10pt}
\end{figure}

\begin{itemize}[leftmargin=*]
  \item  Table~\ref{TableDER} presents the voltage profiles of buses 2 and 3 under different power flow methods. It illustrates the QPF results are identical to the classical results, which validates the correctness of QPF. 
  \item We further study the performance of QPF under a stressed condition, where the loads on buses 3 and 4 are significantly increased to $-1-1j$ p.u. and $-1.6-1j$ p.u.. Fig.~\ref{fig:V_heavy} presents the simulation result. It can be observed that the power flow solution of QPF is still identical to that of the classical method. Although the iteration number increases to 40 because the system is approaching the noise point, QPF still presents satisfactory and comparable convergence performance against classical power flow methods, i.e., the iteration number of QPF is nearly the same as that of the classical methods. 
\end{itemize}

\vspace{-5pt}

\begin{table}[!ht]
  \caption{Voltage profiles with different methods ({\MakeLowercase{p.u.} })}\label{TableDER}
  \vspace{-5pt}
  \centering
  \begin{threeparttable}
\begin{tabular}{p{14mm}<{\centering} p{6mm}<{\raggedright} p{8mm}<{\centering} p{8mm}<{\centering} p{9mm}<{\centering} p{9mm}<{\centering}}
\toprule
   Algorithm & Iteration & $\mathbf V_1$ & $\mathbf V_2$ & $\boldsymbol{\theta}_1$ & $\boldsymbol{\theta}_2$\\
  \midrule 
 \multirow{7}*{QPF} & ~~~1  & 1.0921 & 1.0708 &-0.0041 & -0.0580 \\
   &~~~2  & 1.0821 & 1.0419 & -0.0304 & -0.0738\\
  & ~~~3 & 1.0734 & 1.0357 & -0.0286 & -0.0675 \\
   &~~~4 & 1.0750 & 1.0390 & -0.0264 & -0.0663  \\
  & ~~~5 & 1.0756 & 1.0394 & -0.0268 & -0.0670 \\
    & ~~~6 & 1.0755 & 1.0392 & -0.0268 & -0.0670 \\
  & \cellcolor{green!25}~~~7* & \cellcolor{green!25}1.0755 & \cellcolor{green!25}1.0392 & \cellcolor{green!25}-0.0268 & \cellcolor{green!25}-0.0670 \\
  \midrule
  \multirow{7}*{\makecell{Classical Fast \\ Decoupled}} & ~~~1  & 1.0912 & 1.0686 &-0.0032 & -0.0603 \\
   &~~~2  & 1.0822 & 1.0398 & -0.0305 & -0.0732\\
  & ~~~3 & 1.0733 & 1.0360 & -0.0287 & -0.0667 \\
   &~~~4 & 1.0750 & 1.0395 & -0.0262 & -0.0663  \\
  & ~~~5 & 1.0757 & 1.0394 & -0.0266 & -0.0670 \\
    & ~~~6 & 1.0755 & 1.0391 & -0.0268 & -0.0670 \\
  & \cellcolor{green!25}~~~7* & \cellcolor{green!25}1.0755 & \cellcolor{green!25}1.0392 & \cellcolor{green!25}-0.0268 & \cellcolor{green!25}-0.0670 \\
  \midrule
  \multirow{3}*{\makecell{Classical Newton \\ Raphson}}& ~~~1  & 1.0859 & 1.0479 &-0.0282 & -0.0722 \\
   &~~~2  & 1.0757 & 1.0393 & -0.0268 & -0.0670\\
  & \cellcolor{green!25}~~~3* & \cellcolor{green!25}1.0755 & \cellcolor{green!25}1.0392 & \cellcolor{green!25}-0.0268 & \cellcolor{green!25}-0.0670 \\
  \bottomrule
\end{tabular}
\begin{tablenotes}
\item[*]     Final power flow result
\end{tablenotes}
\end{threeparttable}
\end{table}

\vspace{-5pt}
\subsubsection{Comparison with HHL-QPF} The devised QPF method is more compatible than HHL-QPF~\cite{feng2021quantum} on today's real, noisy quantum devices. For example, in the five-bus test system,  the depth of the compiled circuit optimized from the devised NISQ-QPF algorithm  is 15, which is smaller than that of the HHL-QPF method (i.e., 14811) on the same quantum basis. This indicates the NISQ-QPF method can significantly relieve the impact of noisy quantum environments. 

\begin{figure*}[!ht]
  \centering \includegraphics[width=1\textwidth]{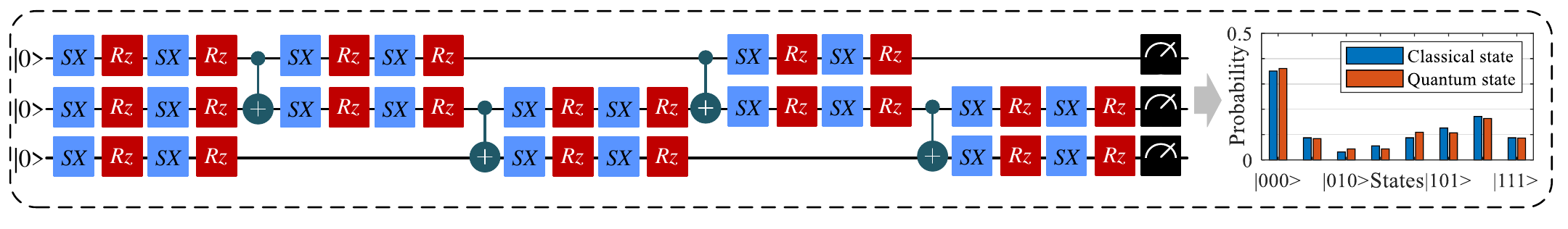}
  \caption{QPF circuit of the 9-bus test system and its output on real quantum machine}
  \label{fig:circuit_9bus}
  \vspace{-10pt}
\end{figure*}

\subsection{NISQ-QPF Test on Noisy Quantum Machine}
This subsection further verifies the performance of NISQ-QPF  using a nine-bus test system on a real IBM quantum hardware {\it{IBMQ\_belem}}. {\it{IBMQ\_belem}} is a 5-qubit quantum computer, whose median CNOT error is $1.235e^{-2}$ and median readout error is $2.85e^{-2}$. 

\subsubsection{Performance of NISQ-QPF on {\it{IBMQ\_belem}}} 
Fig.\ref{fig:circuit_9bus} presents the optimized QPF circuit for a single iteration in the power flow calculation of the nine-bus test system. It can be observed that the NISQ-QPF leads to a shallow circuit depth and fewer CNOT gates. While NISQ-QPF only employs 4 CNOT gates in a 24-depth circuit, HHL-QPF requires a 151795-depth circuit and 54004 CNOT gates for the same system. Therefore, NISQ-QPF is more scalable than HHL-QPF. More importantly, the shallow-depth quantum circuit is significantly more noise-resilient. As shown in Fig.\ref{fig:circuit_9bus}, there only exist slight differences between the measurements from the real quantum hardware and the classical solutions, which indicates NISQ-QPF maintains high performance even on today's noisy machines.

Meanwhile, Table~\ref{TableVQPF} presents the voltage profiles under different methods. It can be observed that QPF results obtained from real quantum hardware are identical to those from the classical methods. Therefore, although noises perturb the quantum solutions, QPF remains to achieve convergence and provide high-fidelity power flow solutions.

\begin{table}
  \caption{Voltage profiles of the 9-bus system with different methods }\label{TableVQPF}
  \vspace{-5pt}
  \centering
  \begin{threeparttable}
\begin{tabular}{ p{10mm}<{\centering} p{10mm}<{\centering} p{10mm}<{\centering} p{10mm}<{\centering} p{10mm}<{\centering}}
\toprule%
    Bus No. & $\mathbf{V}_{quantum}$ & $\mathbf{V}_{classical}$ & $\boldsymbol{\theta}_{quantum}$ & $\boldsymbol{\theta}_{classical}$\\
  \midrule 
 ~~~1  & 1.0204 & 1.0204 & 0.0153 & 0.0153 \\
   ~~~2  & 1.0183 & 1.0183 & 0.0039 & 0.0039\\
   ~~~3 & 1.0277 & 1.0277 & -0.0062 & -0.0062 \\
   ~~~4 & 1.0220 & 1.0220 & -0.0095 & -0.0095  \\
   ~~~5 & 1.0194 & 1.0194 & -0.0042 & -0.0042 \\
     ~~~6 & 1.0178 & 1.0178 & -0.0070 & -0.0070 \\
     ~~~7 & 1.0197 & 1.0197 & -0.0003 & -0.0003 \\
     ~~~8 & 1.0212 & 1.0212 &  -0.0092 & -0.0092 \\
     ~~~9 & 1.0400 & 1.0400 & 0.0000 & 0.0000 \\
  \midrule
  Error(\%) & \multicolumn{2}{c}{0.0035}  & \multicolumn{2}{c}{0.0014}   \\
  \bottomrule
\end{tabular}
\begin{tablenotes}
\item[*]     Final power flow result
\end{tablenotes}
\end{threeparttable}
\vspace{-7pt}
\end{table}

\subsubsection{Extension to stochastic QPF analysis} 
\begin{figure}[!ht]
  \centering
  \includegraphics[width=0.5\textwidth]{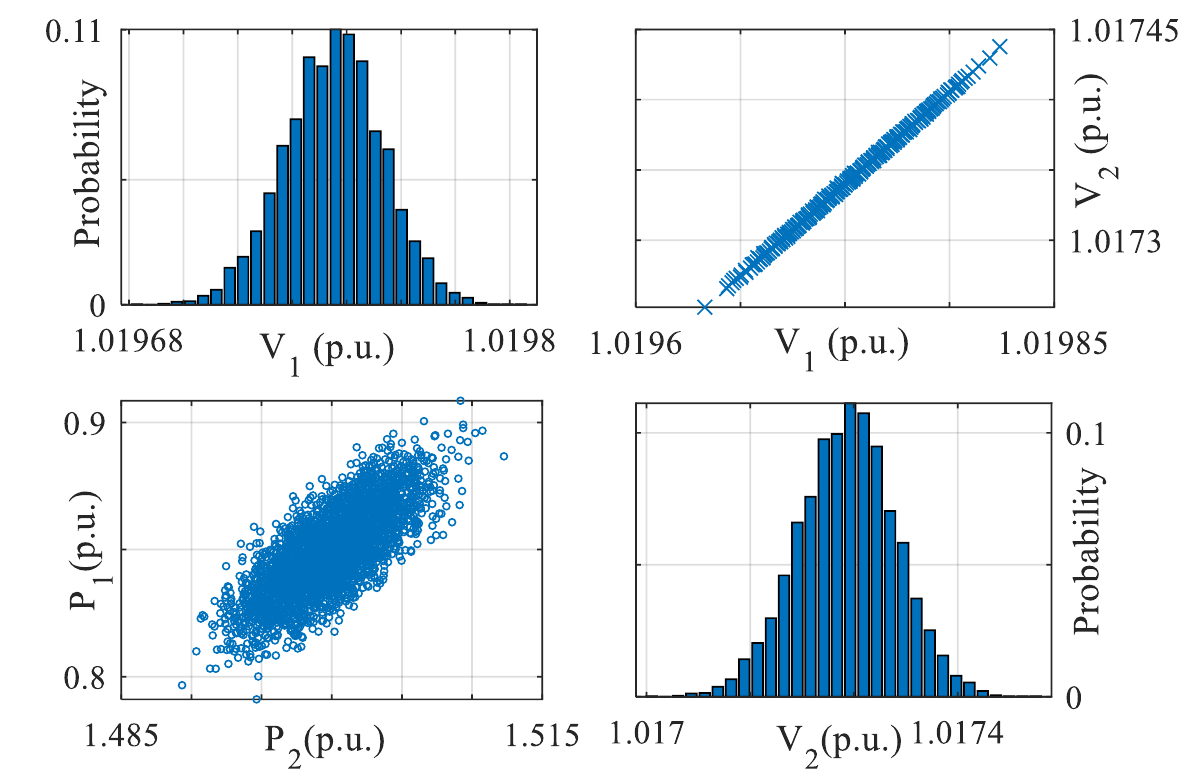}
 \vspace{-0.3cm}
  \caption{Probabilistic voltages at buses 1 and 2 and their correlations { (a) Probabilistic voltages at bus 1 (b) The correlation of voltages between buses 1 and 2 (c) The correlation of power injections between buses 1 and 2 (d) Probabilistic voltages at bus 2.}
  }
  \label{QPF:corelate}
  \vspace{-10pt}
\end{figure}

The devised QPF method is extended to the stochastic power flow analysis. Assuming that power injections at buses 1 and 2 follow correlated Gaussian probability distributions (correlation coefficient: 0.75), 5000 samples are generated stochastically.
Fig.~\ref{QPF:corelate} shows the correlation distributions and probability distributions of voltage magnitudes at buses 1 and 2 obtained from the NISQ-QPF. Therefore,  NISQ-QPF also provides a promising tool for stochastic power flow analysis.
  \vspace{-10pt}
\section{Conclusion}

This paper devises a NISQ-compatible QPF algorithm. Compared with our previous HHL-QPF method, the devised QPF method enables using shallow-depth quantum circuits for power flow calculation, which realizes implementable and reliable power flow analysis on real quantum machines. Case studies validate QPF on a quantum simulator and a real IBM quantum machine {\it{IBMQ\_belem}}. For the future step, we will apply NISQ-QPF on larger grids and verify its scalability.

\bibliographystyle{IEEEtran}
\bibliography{ref}

\end{document}